\documentclass[twocolumn,preprintnumbers,aps,prd]{revtex4}
\usepackage{amssymb,amstext,amsmath,bm,bbm}
\usepackage{xcolor}

\usepackage[normalem]{ulem}

\usepackage{tikzfeynhand}
\usepackage{tikz-cd}
\usepackage{hyperref}
\usetikzlibrary{positioning}
\usetikzlibrary{decorations.markings}

\usepackage[shortcuts]{extdash}

\newcommand{\Dj}{\mbox{\raise0.3ex\hbox{-}\kern-0.4em D}}
\renewcommand{\dj}{d\kern-0.4em\char"16\kern-0.1em}

\newcommand{\ket}[1]{{\vert{#1}\rangle}}
\newcommand{\bbra}[1]{{\langle\!\langle{#1}\vert}}
\newcommand{\kket}[1]{{\vert{#1}\rangle\!\rangle}}

\newcommand{\tr}{\mathop{\rm Tr}\nolimits}

\newcommand{\realni}{\ensuremath{\mathbb{R}}}

\newcommand{\one}{\mathbbm{1}}

\newcommand{\ds}{\displaystyle}

\newcommand{\itGamma}{ {\varGamma} }                                    

\newcommand{\cI}{{\cal I}}

\newcommand{\cM}{{\cal M}}

\newcommand{\cP}{{\cal P}}
\newcommand{\cW}{{\cal W}}

\begin{document}

\title{Challenges for extensions of the process matrix formalism to quantum field theory}

\author{Nikola Paunkovi\'c$^{1}$}
 \email{npaunkov@math.tecnico.ulisboa.pt}

\author{Marko Vojinovi\'c$^{2}$}
 \email{vmarko@ipb.ac.rs}

\affiliation{
$^{1}$Instituto de Telecomunica\~coes and Departamento de Matem\'atica, Instituto Superior T\'ecnico, Universidade de Lisboa, Avenida Rovisco Pais 1, 1049-001, Lisboa, Portugal 
\\
$^{2}$Institute of Physics, University of Belgrade, Pregrevica 118, 11080 Belgrade, Serbia
}


\begin{abstract}
We discuss the issues with tentative generalisations of the process matrix formalism from finite-dimensional mechanical systems all the way to quantum field theory. We present a detailed overview of possible open problems that arise when one attempts to move from particle ontology into the realm of field ontology, i.e., when one transitions from mechanics to field theory framework. These issues need to be addressed, and problems solved, if one aims to expand the scope of applicability of the process matrix formalism, and therefore its usefulness. This is far from a trivial and straightforward endeavour, but rather a task for a whole future research programme.

\end{abstract}

\maketitle

\section{\label{sec:intro}Introduction}

The process matrix formalism has been introduced as a powerful tool to study abstract physical processes, that go beyond standard quantum mecahnics \cite{ore:cos:bru:12}. For example, it enabled the formulation and study of the so-called causal inequalities, which explicitly discuss processes that cannot be explained by a fixed causal order between events \cite{ore:cos:bru:12,bra:ara:fei:cos:bru:15}, as well as other related no-go theorems such as \cite{zyc:cos:pik:bru:17,all:kru:bud:bru:19}. Also, it was extremly useful in the analysis of the properties of the so-called quantum switch protocol \cite{chi:dar:per:val:13,pro:etal:15,rub:roz:fei:ara:zeu:pro:bru:wal:17,rub:roz:mas:ara:zyc:bru:wal:22}. In addition, there are other protocols in quantum information theory where this formalism can be successfully used (for example the analysis of the del Santo-Daki\'c protocol \cite{dak:san:18}, and so on). The purpose of the process matrix formalism is to provide a probability distribution for the outcomes of a very large class of conceptually possible protocols, including those for which we do not know of an experimental realisation \cite{ara:bra:cos:fei:gia:bru:15,bru:15,fei:ara:bru:15,fei:ara:bru:16,ore:gia:16,fei:bru:17,jia:sak:18,cas:gia:Bru:18,gue:rub:bru:19,bav:ara:bru:que:19,pau:voj:20,adl:23}.

The standard formulation of the process matrix formalism is given only for finite-dimensional mechanical systems, such as qubits. Given the mentioned importance of the formalism, one is tempted to generalise it to other mechanical systems as well, and ultimately to field theory. The process matrix formalism can be extended in three main stages. The first extension deals with the input and output spaces which are generalized to the infinite-dimensional separable Hilbert space (such as in the case of a harmonic oscillator). The second extension deals with the nonseparable Hilbert spaces (such as in the case of a free particle). Finally, the third extension deals with the full Fock space, and other properties necessary for the successful description of quantum field theory (QFT) processes.

However, the existing literature that discusses these generalisations is scarce (see for example \cite{fal:pau:voj:23}), since each stage of the above extensions faces nontrivial theoretical problems, of both technical and conceptual nature. The main purpose of our paper is to give an overview of these issues, with an emphasis on the ones opened by field theory. Our analysis suggests that the generalisation of the standard process matrix formalism to QFT is far from a trivial and straightforward endeavour, but rather a task for a whole future research programme.

The paper is organised as follows. In Section~\ref{Sec:ProcessMatrix} we briefly review the process matrix formalism for finite dimensional mechanical systems, and point out possible problems of its extension to the infinite dimensional cases (both separable and nonseparable). In Section~\ref{Sec:BetaDecay} we study beta decay as a prime toy example which serves as an illustration of the problems one faces when trying to generalise the process matrix formalism to QFT. We discuss each of these issues in turn. In Section~\ref{Sec:Conclusions} we give a brief summary and discussion, as well as prospects for future research.

\section{\label{Sec:ProcessMatrix}The process matrix formalism}

The fundamental notions of the process matrix formalism are a set of laboratories and a set of quantum systems being exchanged between them. Each laboratory receives an {\em input} quantum system, and sends an {\em output} quantum system, while inside of a laboratory these two systems can be manipulated by a so-called {\em instrument}, which represets the most general quantum operation that can be performed over the two systems. It is assumed that inside laboratories the usual laws of quantum theory hold, and it is also assumed that the size of each laboratory and the durations of its operations are small enough to be considered negligible for the quantum protocol under consideration. The latter assumption allows one to assign a specific spacetime point to each operation of a given laboratory. This leads one to introduce the notion of a {\em gate}, which represents the action of an instrument at a given spacetime point (see Section~2 of~\cite{pau:voj:20}). In what follows, we will therefore denote both the gate and its corresponding spacetime point by the same symbol, $G$. Also, we denote the input and output Hilbert spaces of the corresponding quantum systems by $G_I$ and $G_O$, respectively. In the standard formulation of the process matrix formalism, these spaces are assumed to be {\em finite-dimensional} \cite{ore:cos:bru:12,ara:bra:cos:fei:gia:bru:15}. The action of an instrument on quatum systems is described as follows. Given some classical input information $a$ and some readout $x$ of an instrument at $G$, the instrument maps the input state $\rho_I$ into an output state $\rho_O = \cM^G_{x,a}(\rho_I)$, where the action of the instrument is described by an operator $\cM^G_{x,a}:G_I\otimes G_I^* \to G_O \otimes G_O^*$.

Within this scenario, one can introduce the notion of a {\em process}, which is represented by a functional over all gates, and denoted $\cW$, so that:
\begin{equation}
\label{eq:ProbabilityFromProcessMatrix}
p(x,y,\dots \vert a,b,\dots) = \cW ( \cM^{G^{(1)}}_{x,a} \otimes \cM^{G^{(2)}}_{y,b} \otimes \dots )\,.
\end{equation}
Here, $p(x,y,\dots \vert a,b,\dots)$ is the probability for obtaining the results $x,y,\dots $, given the inputs $a,b,\dots $. Since the left-hand side is understood to be a probability distribution, the expression on the right-hand side must satisfy the following three axioms,
\begin{equation}
\label{eq:ProcessMatrixAxioms}
\cW \geq 0\,, \quad \tr \cW = \prod_i \dim G^{(i)}_O\,, \quad \cW = \cP_G (\cW)\,,
\end{equation}
where the positivity of the probability distribution is guaranteed by the first axiom, while the normalisation is guaranteed by the second and third axioms. The $\cP_G$ is a particular projector onto a subspace of $\bigotimes_i \left(G^{(i)}_I \otimes G^{(i)}_O\right)$, defined in detail in \cite{ara:bra:cos:fei:gia:bru:15}.

In order to represent an operation $\cM^G_{x,a}$ as a matrix, one can apply the so-called Choi-Jamio\l kowski (CJ) isomorphism that acts over the operations of instruments. This provides us with a computationally more convenient formalism, where the corresponding matrix is defined as
\begin{equation} 
\label{eq:DefCJmap}
M^G_{x,a} = \! \Big[ \! \left( \cI \otimes \cM^G_{x,a} \right) \left( \kket{\one} \bbra{\one} \right) \! \Big]^T \!\!\!\!  \in (G_I\otimes G_O) \otimes (G_I\otimes G_O)^*,
\end{equation}
where
\begin{equation} 
\label{eq:DefTransportVector}
\kket{\one} \equiv \sum_i \ket{i} \otimes \ket{i} \in G_I \otimes G_I
\end{equation}
is a non-normalised maximally entangled state, called the {\em transport vector}, while $\cI$ is the identity. The process $\cW$ is thus given by a {\em process matrix} $W$ that determines the conditional probability (\ref{eq:ProbabilityFromProcessMatrix}) via the equation
\begin{equation} 
\label{eq:VerovatnocaPrekoProcesMatrice}
p(x,y,\dots \vert a,b,\dots) = \tr \left[ ( M^{G_1}_{x,a} \otimes M^{G_2}_{y,b} \otimes \dots ) W \right] \,.
\end{equation}

In certain circumstances, specifically when the instrument $\cM^G_{x,a}$ is linear, one can employ the simpler ``vector'' notation (see Appendix A.1 in~\cite{ara:bra:cos:fei:gia:bru:15}),
\begin{equation} 
\label{eq:DefGateActionVector}
\kket{(\cM^{G}_{x,a})^*} \equiv \left[ \cI \otimes (\cM^{G}_{x,a})^* \right] \kket{\one} \in G_I\otimes G_O\,,
\end{equation}
resulting in
\begin{equation}
\label{eq:DefMatrixPrekoGateVectora}
M^G_{x,a} = \kket{(\cM^{G}_{x,a})^*} \bbra{(\cM^{G}_{x,a})^*}\,.
\end{equation}
Additionally, if the process matrix $W$ is a one-dimensional projector and if all instruments are linear, one can also simplify the expression (\ref{eq:VerovatnocaPrekoProcesMatrice}) for the probability distribution using the notion of a {\em process vector} $\kket{W}$. It is introduced so that $W = \kket{W} \bbra{W}$, which then gives:
\begin{equation} 
\label{eq:DefProbabilityDistribution}
\begin{array}{c}
p(x,y,\dots \vert a,b,\dots) = \vphantom{\ds\int}\hphantom{mmmmmmmmmmmmm}\\
\hphantom{mmmm}\left\Vert \left( \bbra{\cM^{G^{(1)}*}_{x,a}} \otimes \bbra{\cM^{G^{(2)}*}_{y,b}} \otimes \dots \right) \kket{W} \strut \right\Vert^2 . \\
\end{array}
\end{equation}

As described above, the process matrix formalism assumes that all Hilbert spaces involved are finite-dimensional, which can be explicitly seen from the second axiom in (\ref{eq:ProcessMatrixAxioms}), and is also implicitly assumed in the third axiom (the definition of the projector $\cP_G$ involves division with dimensions of input and output spaces, see \cite{ara:bra:cos:fei:gia:bru:15}). Therefore, any provisional generalisation of the process matrix formalism to the framework of QFT faces the inadequacy of this assumption, given that the Fock space is infinite-dimensional and moreover {\em nonseparable}. On one hand, the process matrix axioms explicitly make use of finite dimensions, while on the other hand, the standard rigorous definition of the CJ isomorphism is also establised only for finite-dimensional spaces (see \cite{hol:11a} for the {\em separable} infinite-dimensional case). In the next Section, we will illustrate both the mentioned two issues, and some additional problems which arise in the context of QFT, on the prototype example of beta decay.

\section{\label{Sec:BetaDecay}Process matrix description of beta decay}

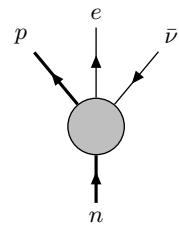
\begin{figure}[!ht]
\centering
\begin{tikzpicture}
\begin{feynhand}
\setlength{\feynhandarrowsize}{4pt}
  \vertex (n) at (0,-0.2) {$n$};
  \vertex [grayblob] (b) at (0,1) {};
  \vertex (p) at (-1,2.2) {$p$};
  \vertex (e) at (0,2.5) {$e$};
  \vertex (nu) at (1,2.2) {$\bar{\nu}$};
\setlength{\feynhandlinesize}{1.2pt}
  \propag[fer] (n) to (b);
  \propag[fer] (b) to (p);
\setlength{\feynhandlinesize}{0.5pt}
  \propag[fer] (b) to (e);
  \propag[antfer] (b) to (nu);
\end{feynhand}
\end{tikzpicture}
\caption{The full Feynman diagram for beta decay.}
\label{SlikaFeynmanovogDijagrama}
\end{figure}

\begin{figure}[!ht]
\centering
\begin{tikzpicture}
\begin{feynhand}
\setlength{\feynhandarrowsize}{4pt}
\propag[fer] (0.5,1) to (0.5,1.5);
\propag[fer] (2,1) -- (2,1.5);
\propag[fer] (3.5,1) -- (3.5,1.5);
\propag[fer] (2,-0.5) -- (2,0);
\filldraw[fill=lightgray] (-0.3,0) rectangle (4.3,1);
\node at (2,0.5) {$W_\beta$};
\node at (2,-0.8) {$M^{(n)}_{t_i}$};
\node at (0.5,1.8) {$M^{(p)}_{t_f}$};
\node at (2,1.8) {$M^{(e)}_{t_f}$};
\node at (3.5,1.8) {$M^{(\bar{\nu})}_{t_f}$};
\draw (1.5,-0.5) -- (2.5,-0.5);
\draw (0,1.5) -- (1,1.5);
\draw (1.5,1.5) -- (2.5,1.5);
\draw (3,1.5) -- (4,1.5);
\propag[plain] (1.5,-0.5) to [out=270, in=180] (2,-1.2);
\propag[plain] (2,-1.2) to [out=0, in=270] (2.5,-0.5);
\propag[plain] (0,1.5) to [out=90, in=180] (0.5,2.2);
\propag[plain] (0.5,2.2) to [out=0, in=90] (1,1.5);
\propag[plain] (1.5,1.5) to [out=90, in=180] (2,2.2);
\propag[plain] (2,2.2) to [out=0, in=90] (2.5,1.5);
\propag[plain] (3,1.5) to [out=90, in=180] (3.5,2.2);
\propag[plain] (3.5,2.2) to [out=0, in=90] (4,1.5);
\end{feynhand}
\end{tikzpicture}
\caption{The process matrix diagram for beta decay.}
\label{SlikaProcesMatrice}
\end{figure}
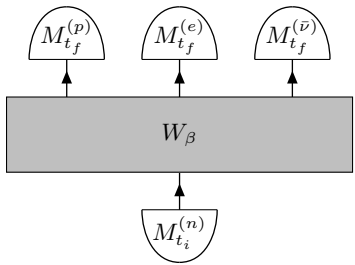

In elementary particle physics, a paradigmatic process that illustrates most of the crucial features of QFT is beta decay. It describes the decay of a neutron into a proton, an electron and an antineutrino:
\begin{equation} \label{eq:BetaDecay}
n \to p^+ + e^- + \bar{\nu}_e \,.
\end{equation}
Using the standard QFT approach, one associates the Feynman diagram to this process as depicted in Figure~\ref{SlikaFeynmanovogDijagrama}.
On the other hand, one can try to describe beta decay using the process matrix formalism, associating to it the process matrix diagram depicted in Figure~\ref{SlikaProcesMatrice}.
The connection between the Feynman's and the process matrix approaches goes beyond the obvious similarity of the two diagrams. Namely, the evaluation of the Feynman diagram gives the amplitude for the process, from which one can obtain the decay width, $\itGamma$, or equivalently the mean lifetime of neutron, $\tau = 1 / \itGamma$. This decay width enters the formula for the number $N(t_f)$ of neutrons in the ensemble at the final time $t_f$,
$$
N(t_f) = N(t_i)\, e^{-\itGamma (t_f-t_i)}\,,
$$
where $N(t_i)$ is the number of neutrons in the ensemble at initial time $t_i$. Therefore, the probability for an individual neutron to decay after time $t_f - t_i$ is given as:
\begin{equation} \label{eq:BetaProb}
p(t_f | t_i) = 1 - \frac{N(t_f)}{N(t_i)} = 1 - e^{-\itGamma (t_f-t_i)}\,.
\end{equation}
However, this probability is precisely the output (\ref{eq:VerovatnocaPrekoProcesMatrice}) of the process matrix description (Figure~\ref{SlikaProcesMatrice}), when applied to beta decay:
\begin{equation} \label{eq:VerovatnocaBeta}
p(t_f | t_i) = \tr \left[ \left( M^{(n)}_{t_i} \otimes M^{(p)}_{t_f} \otimes M^{(e)}_{t_f} \otimes M^{(\bar{\nu}_e)}_{t_f} \right) W_\beta \right] \,.
\end{equation}
This establishes the correspondence (in terms of the desired aim one might even say {\em equivalence}) between the Feynman and the process matrix diagrams, since they both contain the same physical information about the process.

With this in mind, let us now discuss the main issues one faces when attempting to generalise the process matrix formalism to QFT.

\subsection{\label{subsec:nonseparability}Nonseparability of Hilbert spaces}

To begin with, note that while QFT and the process matrix formalism are equivalent at the operational level of predicting the probabilities of the processes, (\ref{eq:BetaProb}) and (\ref{eq:VerovatnocaBeta}), respectively, they do not represent two equivalent mathematical formalisms. Namely, recall that the paradigm of QFT is based on the idea of particle scattering. This assumes the notions of asymptotically free particles as input and output states. It is well known already from QM that the state of a free particle is an eigenvalue of the momentum, and as such is not normalizable. Technically speaking, the Hilbert space of a free particle is not just infinite-dimensional, but also nonseparable, since it contains uncountably infinitely many momentum eigenstates, one for each $\vec{p} \in \realni^3$.

On the other hand, the process matrix formalism is based on the axioms (\ref{eq:ProcessMatrixAxioms}) and the technical notion of CJ isomorphism, both of which are tailored to finite-dimensional Hilbert spaces. Therefore, the first issue is that the axioms explicitly depend on the dimension of the Hilbert space, which is infinite in the case of a free particle. And the second issue, as we have remarked above, is that the CJ isomorphism has not been formulated for nonseparable Hilbert spaces, rendering the case of a free particle an open technical problem.

\subsection{\label{subsec:renormalisation}Renormalisation and regularisation}

The next issue is related to the (non)perturbative formulation of QFT. Given the established operational equivalence with the process matrix description, and the analysis from the previous Subsection, one can argue that providing a precise definition for the axioms (\ref{eq:ProcessMatrixAxioms}) of the process matrix formalism in QFT is at least as hard as providing an exact, nonperturbative definition of QFT itself. Unfortunately, the latter is still unknown, and in fact it represents one of the Millenium Prize Problems~\cite{jaf:06,car:jaf:wil:06}. Lacking the nonperturbative formulation of QFT, the only way to define the axioms in process matrix formalism is perturbatively. This method faces the same issues with divergences, having to encode the equivalents of the regularisation and renormalisation schemes used in the perturbative definition of QFT.

\begin{figure}[!ht]
\centering
\begin{tikzpicture}[baseline=(b.base)]
\begin{feynhand}
\setlength{\feynhandarrowsize}{4pt}
  \vertex (n) at (0,-0.2) {$n$};
  \vertex [grayblob] (b) at (0,1) {};
  \vertex (p) at (-1,2.2) {$p$};
  \vertex (e) at (0,2.5) {$e$};
  \vertex (nu) at (1,2.2) {$\bar{\nu}$};
\setlength{\feynhandlinesize}{1.2pt}
  \propag[fer] (n) to (b);
  \propag[fer] (b) to (p);
\setlength{\feynhandlinesize}{0.5pt}
  \propag[fer] (b) to (e);
  \propag[antfer] (b) to (nu);
\end{feynhand}
\end{tikzpicture}
$=$
\begin{tikzpicture}[baseline=(b.base)]
\begin{feynhand}
\setlength{\feynhandarrowsize}{4pt}
  \vertex (n) at (0,-0.2) {$n$};
  \vertex (b) at (0,1);
  \vertex (c) at (0.5,1.3);
  \vertex (p) at (-1,2.2) {$p$};
  \vertex (e) at (0,2.5) {$e$};
  \vertex (nu) at (1,2.2) {$\bar{\nu}$};
\setlength{\feynhandlinesize}{1.2pt}
  \propag[fer] (n) to (b);
  \propag[fer] (b) to (p);
\setlength{\feynhandlinesize}{0.5pt}
  \propag[bos] (b) to (c);
  \propag[fer] (c) to (e);
  \propag[antfer] (c) to (nu);
\end{feynhand}
\end{tikzpicture}
$+$
\begin{tikzpicture}[baseline=(b.base)]
\begin{feynhand}
\setlength{\feynhandarrowsize}{4pt}
  \vertex (n) at (0,-0.2) {$n$};
  \vertex (bb) at (-0.2,1.5);
  \vertex (b) at (0,1);
  \vertex (cc) at (0.1,1.8);
  \vertex (c) at (0.5,1.3);
  \vertex (p) at (-1,2.2) {$p$};
  \vertex (e) at (0,2.5) {$e$};
  \vertex (nu) at (1,2.2) {$\bar{\nu}$};
\setlength{\feynhandlinesize}{1.2pt}
  \propag[fer] (n) to (b);
  \propag[plain] (b) to (bb);
  \propag[fer] (bb) to (p);
\setlength{\feynhandlinesize}{0.5pt}
  \propag[bos] (b) to (c);
  \propag[bos] (bb) to (cc);
  \propag[plain] (c) to (cc);
  \propag[fer] (cc) to (e);
  \propag[antfer] (c) to (nu);
\end{feynhand}
\end{tikzpicture}
$+\dots +$
\begin{tikzpicture}[baseline=(b.base)]
\begin{feynhand}
\setlength{\feynhandarrowsize}{4pt}
  \vertex (n) at (0,-0.2) {$n$};
  \vertex (b) at (-0.2,1);
  \vertex (ci) at (0.5,1);
  \vertex (cf) at (1.8,1);
  \vertex (cu) at (1.15,1.7);
  \vertex (cd) at (1.15,0.3);
  \vertex (ciu) at (1.15,1.3);
  \vertex (cid) at (1.15,0.7);
  \vertex (bi) at (0,0.5);
  \vertex (bf) at (-0.4,1.5);
  \vertex (c) at (2.5,1);
  \vertex (dl) at (2.3,1.5);
  \vertex (dd) at (2.7,1.5);
  \vertex (sd) at (2.2,1.6);
  \vertex (sl) at (1.5,2.1);
  \vertex (p) at (-1,2.2) {$p$};
  \vertex (e) at (1,2.5) {$e$};
  \vertex (nu) at (3,2.2) {$\bar{\nu}$};
\setlength{\feynhandlinesize}{1.2pt}
  \propag[fer] (n) to (bi);
  \propag[plain] (bi) to (b);
  \propag[plain] (b) to (bf);
  \propag[fer] (bf) to (p);
\setlength{\feynhandlinesize}{0.5pt}
  \propag[glu] (bi) to [out=180, in=220] (bf);
  \propag[fer] (ci) to [out=90, in=180] (cu);
  \propag[plain] (cu) to [out=0, in=90] (cf);
  \propag[fer] (cf) to [out=270, in=0] (cd);
  \propag[plain] (cd) to [out=180, in=270] (ci);
  \propag[bos] (cu) to (ciu);
  \propag[bos] (cd) to (cid);
  \propag[fer] (cid) to [out=0, in=0] (ciu);
  \propag[fer] (ciu) to [out=180, in=180] (cid);
  \propag[bos] (b) to (ci);
  \propag[bos] (cf) to (c);
  \propag[plain] (c) to (dl);
  \propag[plain] (dl) to (sd);
  \propag[fer] (sd) to (sl);
  \propag[plain] (sl) to (e);
  \propag[bos] (sl) to [out=30, in=90] (sd);
  \propag[plain] (c) to (dd);
  \propag[antfer] (dd) to (nu);
  \propag[bos] (dl) to (dd);
\end{feynhand}
\end{tikzpicture}
$+\dots$
\caption{Perturbative expansion of the full Feynman diagram for beta decay.}
\label{RazvojFeynmanovogDijagrama}
\end{figure}
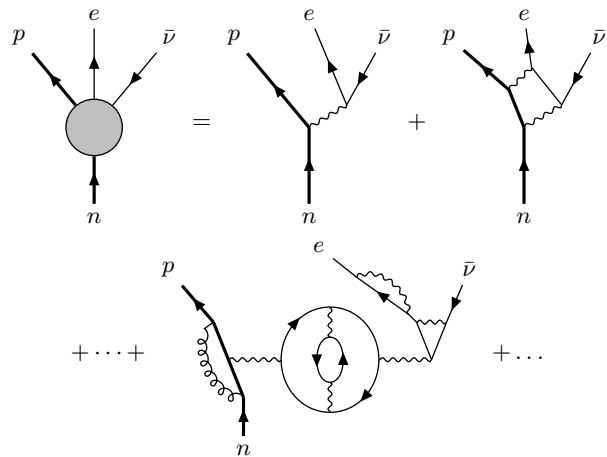

All this can be easily seen at the level of our example. One starts by noting that the Feynman diagram from Figure~\ref{SlikaFeynmanovogDijagrama} represents the full (i.e., nonperturbative) description of beta decay. In practice, the only way to evaluate the diagram is to employ the perturbative expansion, as depicted in Figure~\ref{RazvojFeynmanovogDijagrama}.
Here, the first term is called the tree-level diagram, while all others are called loop diagrams. Although the tree level diagram is straightforward to evaluate and gives a finite amplitude, the loop diagrams are far more intricate, and their evaluation relies on suitable regularisation and renormalisation schemes. The main purpose of these prescriptions is to render the resulting amplitude finite, despite the fact that contributions coming from each particular loop in the diagram are divergent. In fact, one can say that the regularisation and renormalisation schemes actually enter the operational definition of perturbative QFT. As a consequence, any attempt at a perturbative definition of the process matrix formalism will have to mirror these technical issues.

\subsection{\label{subsec:convergence}Radius of convergence}

The perturbative expansion, depicted in Figure~\ref{RazvojFeynmanovogDijagrama}, of the full Feynman diagram from Figure~\ref{SlikaFeynmanovogDijagrama} also suffers from an additional issue. Namely, even after eliminating the divergences of individual diagrams, the expansion itself fails to converge. In fact, perturbative QFT is known to be an asymptotic theory, since at any order $n$ of the expansion, there are approximately $n!$ Feynman diagrams of that order, while the contribution of the perturbation parameter $g$ is $g^{-n}$, i.e., the total amplitude $\cM_n$ at order $n$ is proportional to
$$
\cM_n \sim \frac{n!}{g^n}\,.
$$
Since the factorial grows faster than any power, there will be a finite order $n_c$ after which the series will begin to increase without bound. In other words, the perturbative expansion of a QFT has zero radius of convergence, and thus represents an asymptotic series, rather than a convergent one \cite{dys:52,wei:96}.

Therefore, even if one adopts a perturbative approach to process matrix formalism in QFT, and incorporates certain regularisation and renormalisation schemes in the axioms (\ref{eq:ProcessMatrixAxioms}), it is to be expected that the resulting axioms would be adequate only up to a certain perturbation order. Since in such approach one can only obtain a finite approximation of the exact result, the perturbative probability distribution is in fact not supposed to be normalised. This is especially problematic for the latter two axioms in (\ref{eq:ProcessMatrixAxioms}), since they ensure the normalisation of the probability distribution.

\subsection{\label{subsec:variability}Variability of the number of systems}

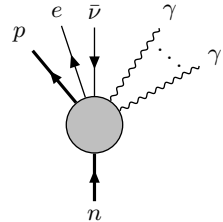
\begin{figure}[!ht]
\centering
\begin{tikzpicture}
\begin{feynhand}
\setlength{\feynhandarrowsize}{4pt}
  \vertex (n) at (0,-0.2) {$n$};
  \vertex [grayblob] (b) at (0,1) {};
  \vertex (p) at (-1,2.2) {$p$};
  \vertex (e) at (-0.5,2.5) {$e$};
  \vertex (nu) at (0,2.5) {$\bar{\nu}$};
  \vertex (gl) at (1,2.5) {$\gamma$};
  \vertex (t) at (1,2) {$\ddots$};
  \vertex (gd) at (1.6,1.9) {$\gamma$};
\setlength{\feynhandlinesize}{1.2pt}
  \propag[fer] (n) to (b);
  \propag[fer] (b) to (p);
\setlength{\feynhandlinesize}{0.5pt}
  \propag[fer] (b) to (e);
  \propag[antfer] (b) to (nu);
  \propag[bos] (b) to (gl);
  \propag[bos] (b) to (gd);
\end{feynhand}
\end{tikzpicture}
\caption{The full Feynman diagram for the radiative beta decay.}
\label{SlikaRadFeynmanovogDijagrama}
\end{figure}

The issues with the formulation of the process matrix framework based on the perturbative QFT do not end here. Namely, looking again at our example of beta decay, let us recall that the neutron can decay in more than one way. One example is the radiative beta decay, given by the following family of processes,
\begin{equation} \label{eq:BetaDecayAlt}
n \to p^+ + e^- + \bar{\nu}_e + k\gamma \,,
\end{equation}
where the last term represents $k$ outgoing photons. This is described by the full Feynman diagram in Figure~\ref{SlikaRadFeynmanovogDijagrama} (for such processes and their perturbative evaluation, see for example \cite{iva:17}).
For each $k$, (\ref{eq:BetaDecayAlt}) specifies one possible decay channel of a neutron, with its own decay width $\itGamma_k$ and the corresponding decay probability $p_k(t_f|t_i)$. Then, the {\em total} decay probability of the neutron via the channels (\ref{eq:BetaDecay}) and (\ref{eq:BetaDecayAlt}) is in fact given via the equation
\begin{equation} \label{eq:TotalBetaProb}
p_{\rm tot}(t_f | t_i) = 1 - e^{-\itGamma_{tot} (t_f - t_i)}\,,
\end{equation}
where $ \itGamma_{\rm tot} = \itGamma + \sum_k \itGamma_k $. In the process matrix formalism, each decay channel is described by a process matrix $W_k$ and the corresponding set of gate operations $M_k$, such that the probability of neutron decay via that particular channel is given by the expression analogous to (\ref{eq:VerovatnocaBeta}). However, for different values of $k$ the {\em number of output systems} and the corresponding gates is not the same, which implies that the corresponding process matrices $W_k$ act on different Hilbert spaces. Therefore, one cannot describe the total probability (\ref{eq:TotalBetaProb}) for beta decay by an equation of the form (\ref{eq:VerovatnocaBeta}), since the latter features a fixed number of gates.

It is not obvious how one should generalise the equation for the total probability $p_{\rm tot}(t_f | t_i)$ in the process matrix formalism to include a variable number of gate operations. Namely, employing (\ref{eq:TotalBetaProb}) and the analog of (\ref{eq:BetaProb}) for each channel separately, one can express the total probability in terms of single-channel ones as follows,
\begin{equation} \label{eq:nonlinearProb}
p_{\rm tot}(t_f | t_i) = 1 - \prod_{k\geq 0} \Big[1-p_k(t_f | t_i)\Big] \,,
\end{equation}
where $p_0(t_f | t_i) \equiv p(t_f | t_i)$ corresponds to the original nonradiative channel (\ref{eq:BetaDecay}). This expression is nonlinear in single-channel probabilities, which means that the process matrix formula for the total probability cannot have the form analogous to (\ref{eq:VerovatnocaPrekoProcesMatrice}), i.e., the process functional $\cW$ {\em cannot be linear} anymore. Therefore, the issues with generalising the process matrix formalism to QFT lie not just in the precise definition of the axioms (\ref{eq:ProcessMatrixAxioms}), but also in the precise definition of the expression for the probability~(\ref{eq:VerovatnocaPrekoProcesMatrice}).

An alternative viewpoint could be to treat all output gates of the process matrix corresponding to the process (\ref{eq:BetaDecayAlt}) as a single big gate. In this scenario, the whole process matrix describes just a simple channel, between one input and one output gate. However, in this case the state of the physical system in the output gate must belong to a full Fock space, since it describes multiple particles, and moreover a variable number of them. Having a process matrix formalism defined over a full Fock space suffers the same technical issues of nonseparability discussed in Subsection \ref{subsec:nonseparability}, since the Fock space is defined as an infinite orthogonal sum of tensor products of Hilbert spaces for a single free particle. Apart from nonseparability, there may be various additional technical issues of limits and convergence due to the fact that the orthogonal sum is infinite. It should be emphasised that by merely replacing a variable number of output gates acting over single-particle Hilbert spaces by a single gate acting over a Fock space, one still does not resolve the above mentioned problem of nonlinearity of the process functional $\cW$ and consequently the resulting output probabilities (\ref{eq:nonlinearProb}).

\subsection{\label{subsec:unruh}Noninertial motion and the Unruh effect}

The problem of variable number of input and output systems manifests itself also in an additional manner, which is not directly a consequence of the established correspondence between Feynman's and process matrix formalisms, but is a generic phenomenon that must be addressed whenever one discusses QFT. Namely, originally the process matrix formalism has been formulated for mechanical systems. When moving from the particle ontology to the field ontology framework of QFT, one of the novel effects that must be accounted for is the Unruh effect (more precisely the Fulling-Davies-Unruh effect \cite{ful:73,dav:75,unr:76}). In short, the Unruh effect says that the same physical state of a system can be interpreted, for example, as a vacuum state by an inertial detector, and as a thermal state by an accelerating detector. In the context of the process matrix formalism, this means that a given gate will receive different input quantum states and transmit different output quantum states, depending on the state of motion of the gate and its corresponding instrument. In other words, the number of input and output systems depends on the state of motion of the gate. All our discussion so far in this paper has implicitly assumed that all gates are inertial. However, in a general situation, noninertial gates should also be a part of the formalism, which means that any given gate should additionally be characterised by its local acceleration vector. This additional information about a {\em state of motion} must therefore enter the fundamental description of the process matrix formalism for~QFT.

\section{\label{Sec:Conclusions}Conclusions}

In this paper, we have discussed the issues and problems that arise when one attempts to generalise the process matrix formalism to the level of QFT, on the standard toy example of beta decay.

In order to better isolate the problems one faces in such generalisation, we have established the correspondence between the Feynman's and the process matrix diagrams, and their resulting probability distributions. Due to this correspondence, we have discussed the following list of issues:
\begin{itemize}
\item It is necessary to work with infinite-dimensional and nonseparable Hilbert spaces, which are incompatible with the standard axioms of the process matrix formalism, as well as with the CJ isomorphism used throughout the formalism (Subsection \ref{subsec:nonseparability}). 
\item Lacking the full nonperturbative formulation of QFT, the only way to define the axioms in process matrix formalism is perturbatively. However, such a perturbative formulation inherits the standard issues of pertubative QFT such as regularisation and renormalisation prescriptions, in order to eliminate infinities from the theory (Subsection \ref{subsec:renormalisation}).
\item Even if one adopts a perturbative approach to process matrix formalism in QFT, and incorporates certain regularisation and renormalisation schemes in the axioms, it is to be expected that the resulting axioms would be adequate only up to a certain perturbation order, since the radius of convergence of the perturbative expansion in QFT is zero, and the theory is asymptotic rather than convergent (Subsection \ref{subsec:convergence}).
\item In QFT there are many processes (inlcuding the beta decay) which feature different channels with variable number of input and output systems. One way to tackle this problem is to move from single-particle Hilbert spaces to Fock spaces, but this also suffers from known and potential new problems related to separability and infinite number of dimensions. Additionally, the variable number of input and output systems implies the  nonlinearity of the process functional and therefore its total probability distribution (Subsection \ref{subsec:variability}).
\item While not directly a consequence of the above mentioned correspondence, one must take into account the state of motion of the gates, whenever one deals with QFT. This is necessary because of the Unruh effect, which predicts different number of systems interacting with an apparatus, depending on the state of motion of the apparatus. This state of motion must enter the the fundamental description of the process matrix formalism for QFT (Subsection~\ref{subsec:unruh}).
\end{itemize}

Finally, note that the issues raised and analysed on the example of beta decay are in fact generic for all interactions between particles in QFT, and as such need to be addressed for the process matrix generalisation discussed here. In conclusion, our analysis suggests that the generalisation of the standard process matrix formalism to QFT is far from a trivial and straightforward endeavour, but rather a task for a whole future research programme.

\bigskip

\centerline{\bf Acknowledgments}

\bigskip

The authors wish to thank \v Caslav Brukner, Fabio Costa, Ricardo Faleiro and Bruno Mera for useful discussions.

NP acknowledges Funda\c{c}\~ao para a Ci\^{e}ncia e Tecnologia (FCT), Instituto de Telecomunica\c{c}\~oes Research Unit, ref. UIDB/50008/2020, UIDP/50008/2020, the European Regional Development Fund (FEDER), through the Competitiveness and Internationalization Operational Programme (COMPETE 2020), under the project QuantumPrime reference: PTDC/EEI-TEL/8017/2020 and QuRUNNER, QUESTS action of Instituto de Telecomunica\c{c}\~oes and the QuantaGENOMICS project, through the EU H2020 QuantERA II Programme, Grant Agreement No 101017733, as well as the FCT Est\'{i}mulo ao Emprego Cient\'{i}fico grant no. CEECIND/04594/2017/CP1393/CT000.


MV was supported by the Ministry of Education, Science and Technological Development of the Republic of Serbia, and by the Science Fund of the Republic of Serbia, grant 7745968, ``Quantum Gravity from Higher Gauge Theory 2021'' --- QGHG-2021. The contents of this publication are the sole responsibility of the authors and can in no way be taken to reflect the views of the Science Fund of the Republic of Serbia.

\onecolumngrid

\bibliography{vac-proc-mat}

\end{document}